# Enabling Multi-Shell b-Value Generalizability of Data-Driven Diffusion Models with Deep SHORE

Vishwesh Nath[1][0000-0002-6840-6205], Ilwoo Lyu[1], Kurt G. Schilling[2], Prasanna Parvathaneni[3], Colin B. Hansen[1], Yucheng Tang[1], Yuankai Huo[1], Vaibhav A. Janve[2], Yurui Gao[2], Iwona Stepniewska[4], Adam W. Anderson[2], Bennett A. Landman[2]

[1] Computer Science, Vanderbilt University, Nashville TN 37203, USA
[2] Biomedical Engineering, Vanderbilt University, Nashville, TN 37203, USA
[3] Electrical Engineering, Vanderbilt University, Nashville, TN
[4] Psychiatry, Vanderbilt University, Nashville, TN
vishwesh.nath@vanderbilt.edu

**Abstract.** Intra-voxel models of the diffusion signal are essential for interpreting organization of the tissue environment at micrometer level with data at millimeter resolution. Recent advances in data driven methods have enabled direct comparison and optimization of methods for *in-vivo* data with externally validated histological sections with both 2-D and 3-D histology. Yet, all existing methods make limiting assumptions of either (1) model-based linkages between b-values or (2) limited associations with single shell data. We generalize prior deep learning models that used single shell spherical harmonic transforms to integrate the recently developed simple harmonic oscillator reconstruction (SHORE) basis. To enable learning on the SHORE manifold, we present an alternative formulation of the fiber orientation distribution (FOD) object using the SHORE basis while representing the observed diffusion weighted data in the SHORE basis. To ensure consistency of hyper-parameter optimization for SHORE, we present our Deep SHORE approach to learn on a data-optimized manifold. Deep SHORE is evaluated with eight-fold cross-validation of a preclinical MRI-histology data with four b-values. Generalizability of *in-vivo* human data is evaluated on two separate 3T MRI scanners. Specificity in terms of angular correlation (ACC) with the preclinical data improved on single shell: 0.78 relative to 0.73 and 0.73, multi-shell: 0.80 relative to 0.74 (p < 0.001). In the *in-vivo* human data, Deep SHORE was more consistent across scanners with 0.63 relative to other multi-shell methods 0.39, 0.52 and 0.57 in terms of ACC. In conclusion, Deep SHORE is a promising method to enable data driven learning with DW-MRI under conditions with varying b-values, number of diffusion shells, and gradient directions per shell.

**Keywords:** Deep Learning, Orthonormal Basis, SHORE, Spherical Harmonics.

## 1    Introduction

Diffusion-weighted magnetic resonance imaging (DW-MRI) is essential for non-invasive reconstruction of the microstructure for the human in-vivo brain. These images are sensitized to the underlying organization of the tissue at a millimetric scale. Multiple approaches have been proposed that can model the non-linear relationship between the DW-MRI signal and biological microstructure with the most common being diffusion tensor imaging (DTI) [1]. Substantial efforts have shown that other advanced approaches can recover more elaborate reconstruction of the microstructure and these methods are collectively referred to as high angular resolution diffusion imaging (HARDI) [2]. HARDI methods have been broadly proposed in two categories of single shell acquisitions and multi-shell acquisitions (i.e., using multiple diffusivity values). A majority of single shell HARDI methods utilize spherical harmonics (SH) based modelling as in q-ball imaging (QBI) [3], super-resolved constrained deconvolution (sCSD) [4], and many others. However, SH based modelling cannot directly leverage additional information provided by multi-shell acquisitions. SH have been combined with other bases to represent multi-shell data, e.g., solid harmonics [5], simple harmonic oscillator reconstruction (SHORE) [6], and spherical polar Fourier imaging [7].

Methodological exploration has been driven through classical mathematical transforms while data-driven approaches have been limited (Fig. 1) due to lack of external validation data. Prior work using data-driven approaches for DW-MRI has been shown in [8], however the primary application for their work is shown for outlier detection and low rank signal prediction. Validation through histology is critical to evaluate the precision of white matter (WM) reconstruction [9]. Prior work through machine learning approaches on reconstruction for single shell diffusion acquisitions has exhibited higher precision and reproducibility. However, this has not been shown for multi-shell acquisitions due to lack of external validation data [10] (Fig 1).

To overcome these issues, we propose a novel approach, Deep SHORE, which incorporates the following key contributions: (1) an unsupervised hyper-parameter optimization for improved learning in the SHORE manifold, (2) representation of a microstructure model in the SHORE manifold to improve precision and reproducibility, and (3) a non-negativity constraint implementation for a deep learning model.

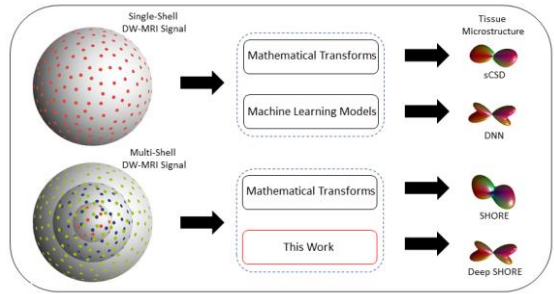

**Fig. 1.** Different classes of methods have been used to infer tissue microstructure from single shell and multi shell DW-MRI data. The gap addressed herein is in data-driven machine learning models for multi-shell DW-MRI data.

## 2 Data Acquisition

Three *ex-vivo* squirrel monkey brains were imaged on a Varian 9.4T scanner. A total of 100 gradient volumes were acquired using a diffusion-weighted EPI sequence at diffusivity values of 3000, 6000, 9000 and 12000 s/mm$^2$ at an isotropic resolution of 0.3mm. An observation is that approximation of b-values for the *ex-vivo* acquisition is equivalent to *in-vivo* b-values of 1000, 2000, 3000 and 4000 s/mm$^2$ [11]. After acquisition, the tissue was sectioned and stained with fluorescent dil and imaged on a LSM710 confocal microscope following procedures outlined in [9]. The histological fiber orientation distribution (HFOD) was extracted using 3D structure tensor analysis. A multi-step registration procedure was used to determine the corresponding diffusion MRI signal [9]. A similar procedure is outlined in [12]. A total of 567 histological voxels were processed. A hundred random rotations were applied to the remaining voxels for both the MR signal and the HFOD to augment the data, bringing the total to 57,267 voxels. As a limitation we acknowledge that there is a possibility of registration error approximately up to the size of MR voxels (up to 300 micrometers) [13].

The *in-vivo* acquisitions of the three human subjects were acquired on two sites 'A' and 'B'. Both sites were equipped with a 3T scanner with a 32-channel head coil. Structural T1 MPRAGE was acquired for all subjects on both the sites. The diffusion acquisition protocol and scanner information are listed on each site as follows:

Site 'A': The scan was acquired at a diffusivity values of 1000, 1500, 2000, 2500, 3000 s/mm$^2$. A total of 96 diffusion weighted gradient volumes were acquired per diffusivity value with a 'b0'. Briefly the other parameters are: SENSE=2.5, partial Fourier=0.77, FOV=96x96, Slice=48, isotropic resolution: 2.5mm.

Site 'B': All parameters of scan acquisition were same as that of the scanner at site 'A' except for the isotropic resolution which was 1.9x1.9x2.5mm$^3$ and down-sampled to 2.5mm iso.

The *in-vivo* acquisitions were pre-processed with standard procedures of eddy, topup and b0 normalization followed by pairwise registration per subject [14]. T1s were registered and transformed to the diffusion space. Brain extraction tool was used for skull stripping [15]. WM segmentation was performed using T1 for *in-vivo* data [16].

## 3 Methods

The SHORE basis function has been shown to capture the representation of multi-shell DW-MRI with minimal representation error [6] and ensure the same when modelling single shell DW-MRI. The DW-MRI normalized signal, *E(q)*, can be represented as:

$$E(q) = \sum_{n=0}^{N} \sum_{l=0}^{n} \sum_{m=-l}^{l} c_{nlm} G_{nl}(q,\zeta) Y_l^m(u) \quad \ldots \quad (1)$$

where *c* are the coefficients to be estimated, *G* depicts the radial basis combined with *Y*, the SH basis. The radial basis *G* is represented as follows:

$$G_{nl}(q,\zeta) = \kappa_{nl}(\zeta)\left(\frac{q^2}{\zeta}\right)^{\frac{l}{2}} \exp\left(-\frac{q^2}{2\zeta}\right) L_{n-\frac{l}{2}}^{l+\frac{1}{2}}\left(\frac{q^2}{\zeta}\right) \quad \ldots \quad (2)$$

where ζ is the scale parameter, *q* is the radius of the diffusivity value, and *L* depicts the associated Laguerre polynomial. Eq (2) can be optimized using the BFGS [17] algorithm by iterative refitting of the coefficients *c*. BFGS is well-known for solving unconstrained non-linear optimization problems. The novelty that we introduce here is that, when functioning across several normalized datasets, an optimal ζ per dataset, will lead to learning on an optimized manifold. Additional parameters of SHORE include: radial order: 6, and regularization constants: 1e-8 [6]. SHORE estimates 50 coefficients at $6^{th}$ order. Regularized linear least squares were used for the estimation of the coefficients. As one notes, l is even for diffusion spherical harmonics or SHORE due to symmetry of diffusion inference process. Essentially, SH at $8^{th}$ order and SHORE at $6^{th}$ order offer the same degree of freedom. SHORE at higher orders is known to suffer from overfitting effects [18].

The HFOD represented as SH coefficients can be fitted to the SHORE basis with the two considerations of (1) diffusivity value and (2) 'ζ' scaling parameter. First the SH coefficients from an $8^{th}$ order were sampled over a sphere of 100 gradient directions. The directions were ensured to be uniformly sampled on the sphere with minimized electrostatic repulsion. These directions were kept consistent at all times while predicting from the network as well. The diffusivity value was set to 2000 s/mm$^2$. After which, it was fitted to SHORE basis using the process described above.

**Non-Negativity.** The FOD, when modelled as SH, cannot exist with negative mean. If it does, then the microstructure exists in the imaginary part of the SH which does hold true when modelling with real even ordered SH. Hence, there was an existing gap to enforce non-negativity on a deep learning network while training and making predictions. We use a regularization value of 0.005 to truncate all values on a set of gradient directions where value is <=0 on both sides of input and output. Thereafter, log space is used instead of linear space: *ln(E(q))* and *ln(P(r))*, where E(q) is the normalized signal and P(r) is the FOD sampled over the gradient directions. Fitting of the representation method such as SH or SHORE follows after log transformation. After the predictions are made, the coefficients of a representation are transformed using exponential to recover them back to linear space.

**Deep Network Design.** We use a 5-layered deep network with the following number of neurons: x1:400, x2:45, x3:200, x4:45 and x5:200. A residual block was created for the layers x2, x3 and x4 and hence the number of neurons was kept equal for x2 and x4. All layers were activated with 'elu'. Additional parameters of the network: Loss function: mean squared error, batch size: 1000, optimizer: RMSProp. While training, only the input and output coefficients were modified for different subcases (discussed in next section). This was due to the fact that SHORE at $6^{th}$ radial order is defined by 50 coefficients and SH at $8^{th}$ order is defined by 45 coefficients. For training of the network, we used k-fold cross validation where k=5 for optimal training.

To evaluate on any withheld set of data we used the angular correlation coefficient (ACC) [19], which is a measure on a scale of -1 to 1 where 1 is the best correlation. ACC is defined using two sets of SH coefficients 'u' and 'v':

$$ACC = \frac{\sum_{j=1}^{\infty} \sum_{m=-j}^{j} u_{jm} v_{jm}^*}{\left[\sum_{j=1}^{\infty} \sum_{m=-j}^{j} |u_{jm}|^2\right]^{0.5} \cdot \left[\sum_{j=1}^{\infty} \sum_{m=-j}^{j} |v_{jm}|^2\right]^{0.5}} \quad \ldots \quad (3)$$

**Evaluation Strategies.** From the total of 57,267, we create 8 testing sets of data where each set has 7,272 voxels except for the last one which has 6,363 voxels. The remaining data for each set were used as training. For all cross-validation experiments, blocks of 101 voxels were randomly allocated to testing/training cohorts ensuring that no synthetic rotations of training data were included in the testing phase. While fitting SHORE coefficients the diffusivity shell of 6000 s/mm$^2$ was withheld leading to four cases of evaluation in incrementing order of shells. For evaluation purposes, we used three sub-cases of deep learning manifolds 1.) Input of '$\zeta$' optimized SHORE DW-MRI and output SH-HFOD. 2) Input of unoptimized '$\zeta$' SHORE-DWMRI and output of SHORE-HFOD 3.) Input of optimized '$\zeta$' SHORE-DWMRI and output of SHORE-HFOD.

Furthermore, we make comparisons between single-shell and multi-shell approaches. For single shell, we show the comparison between the leading single shell approach sCSD, a prior proposed approach that utilizes deep learning [10, 20] (SHDNN) and SHORE derived FOD on the withheld shell and the same for multi-shell where sCSD and SHDNN were excluded.

For *in-vivo* reproducibility evaluation, we compare the ACC for all the pairs of WM voxels between the two sites 'A' and 'B' on a per subject basis. For SHORE based approaches, all the five shells of data were used while for sCSD we used b-value of 2000 s/mm$^2$ as the highest reproducibility was exhibited on the specific shell.

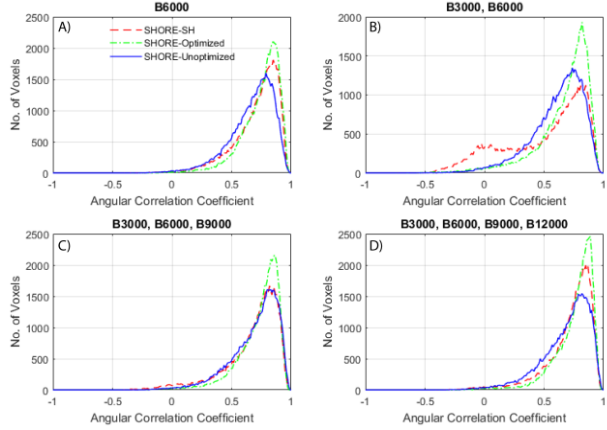

**Fig. 2.** ACC of paired voxels between HFOD and predictions of the deep learning methods across three different manifolds (1) Optimized SHORE -> SH (2) Optimized SHORE -> SHORE (3) Unoptimized SHORE -> SHORE across all 57,267 voxels. Predictions across A) single shell of b-value 6000 s/mm$^2$. B) Two shells of 3000 and 6000 s/mm$^2$. C) Three shells of 3000, 6000 and 9000 s/mm$^2$. D) Four shells of 3000 -12000 s/mm$^2$.

## 4    Results

Evaluation across the withheld shell using combinations with other shells shows that when learning across the optimized SHORE manifold the ACC is most skewed towards higher correlation as compared to the other two approaches (Fig 2). The median of all distributions for each method is presented in Table 1. The median for optimized SHORE learning is the highest. We found significant improvements after non-parametric signed rank test for all pairs of distributions (p<<0.001, Wilcoxon signed rank test).

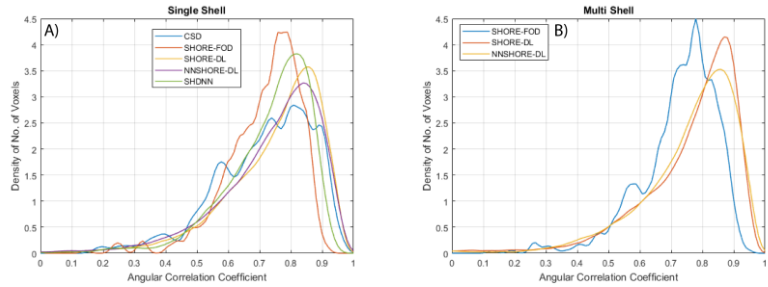

**Fig. 3.** A) Comparison of single shell approaches on the diffusivity shell of 6000 s/mm$^2$ using ACC on all pairs of voxels of predictions of different methods with HFOD. B) Comparison of multi-shell approaches on all four shells between 3000 – 12000 s/mm$^2$ using ACC on all pairs of voxels of predictions for different methods.

**Table 1.** Median & mean values of 4 dataset combinations for the deep learning approaches.

| Deep Learning Approaches | One Shell | | Two Shell | | Three Shell | | Four Shell | |
|---|---|---|---|---|---|---|---|---|
| | Median | Mean | Median | Mean | Median | Mean | Median | Mean |
| U SHORE → SHORE HFOD | 0.70 | 0.67 | 0.65 | 0.61 | 0.74 | 0.69 | 0.73 | 0.68 |
| O SHORE → SH HFOD | 0.75 | 0.71 | 0.61 | 0.50 | 0.74 | 0.67 | 0.77 | 0.72 |
| O SHORE → SHORE HFOD | **0.78** | **0.75** | **0.73** | **0.67** | **0.77** | **0.73** | **0.80** | **0.76** |

U-Unoptimized, O-Optimized. All methods were intercompared per shell combination, using Wilcoxon signed rank test and corrected using Bonferroni correction. All combinations were found to be significant.

When comparing predictions across single shell methods (Fig 3A), the trend in the following increasing order of correlation (median in parenthesis): sCSD (0.73), SHORE-FOD (0.74), SHDNN (0.76), NNSHORE-DL (0.77), SHORE-DL (0.78). Similarly, when making multi-shell comparisons (Fig 3B) we can observe increasing order of correlation: SHORE-FOD (0.75), NNSHORE-DL (0.79) and SHORE-DL (0.80). Non-parametric signed rank test for all pairs of distributions were found to be p << 0.001.

Observing the distribution of ACC across all pairs of WM voxels per subject (Fig 4), increasing level of reproducibility from sCSD, SHORE-DL, SHORE-FOD and NNSHORE-DL. Non-parametric signed rank test for all pairs of distributions were found to be p << 0.001 (Table 2). The NNSHORE-DL exhibits highest reproducibility across all three subjects.

**Table 2.** Median and mean values of ACC for WM voxels across 3 subjects for the methods.

| Method | Subject 1 | | Subject 2 | | Subject 3 | |
| --- | --- | --- | --- | --- | --- | --- |
| | Median | Mean | Median | Mean | Median | Mean |
| Super resolved CSD | 0.37 | 0.37 | 0.49 | 0.47 | 0.31 | 0.31 |
| SHORE-DL | 0.49 | 0.41 | 0.59 | 0.49 | 0.48 | 0.42 |
| SHORE-FOD | 0.59 | 0.50 | 0.64 | 0.53 | 0.49 | 0.42 |
| NNSHORE-DL | **0.63** | **0.52** | **0.67** | **0.56** | **0.61** | **0.52** |

All methods were intercompared per subject, using Wilcoxon signed rank test and corrected using Bonferroni correction. All combinations were found to be significant.

Qualitatively we can observe that SHORE-FOD exhibits higher reproducibility than sCSD and NNSHORE-DL exhibits higher as compared to SHORE-FOD (Fig 5).

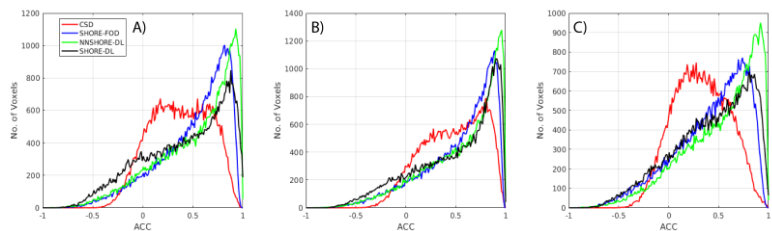

**Fig. 4.** Comparison of proposed approaches with baselines of sCSD and SHORE-FOD across all pairs of WM voxels between the scans of site 'A' and 'B' for each subject. A) Subject 1 B) Subject 2 and C) Subject 3

## 5  Conclusion

Deep SHORE is the first data-driven approach that generalizes diffusion microstructure estimation across multiple b-values, radial b-value sampling, and angular orientation sampling. Our approach enables direct comparison of data-driven diffusion analyses with model-based methods, e.g., sCSD, SHORE-FOD. Although Deep SHORE (NNSHORE-DL) compares favorably when subjected to quantitative cross-validation against histology data (Figs. 3 & 4, Tables 1 & 2), the total amount of data available is a limitation of this study. As current and planned studies acquire more data, we will be able to better train and evaluate data-driven approaches.

**Acknowledgements:** NIH R01EB017230, NIH R01NS058639, NIH T32EB001628, 1S10OD020154-01, UL1 RR024975-01, UL1 TR000445-06 and Nvidia for GPU hardware.

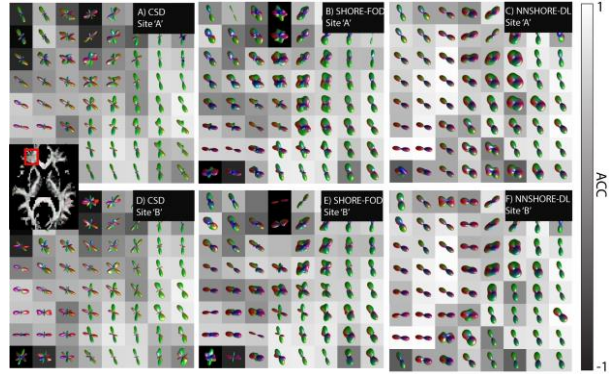

**Fig. 5.** We focus on the ROI of left side frontal lobe of WM. The glyphs depict the FOD's derived from baselines of sCSD, SHORE-FOD and the proposed NNSHORE-DL. The underlay depicts the scalar measure of ACC. A & D) sCSD reconstructed on site 'A' and 'B'. B & E) SHORE-FOD reconstructed on site 'A' and 'B'. C & F) NNSHORE-DL reconstructed on site 'A' and 'B'.